\documentclass[12pt]{article}
\usepackage{latexsym} 
\usepackage{epsfig}
\title{A search for ferromagnetism 
in transition-metal-doped piezoelectric ZnO}

\author{Nicola A. Spaldin\\
Materials Department, University of California, \\
Santa Barbara, CA 93106-5050}

\begin{document}

\maketitle

\begin{abstract}

We present the results of a computational study of ZnO in the presence of
Co and Mn substitutional impurities.
The goal of our work is to identify potential ferromagnetic ground states
within the (Zn,Co)O or (Zn,Mn)O material systems that are also good candidates 
for piezoelectricity. We find that, in contrast to previous results, robust
ferromagnetism is not obtained by substitution of Co or Mn on the Zn site,
unless additional carriers (holes) are also incorporated.
We propose a practical scheme for achieving such $p$-type
doping in ZnO.

\end{abstract}

\newpage
\section{Introduction}

The observation of ferromagnetism in diluted magnetic
semiconductors such as (Ga,Mn)As\cite{Ohno} has spawned a great
deal of recent research in the field now popularly
known as ``spintronics''\cite{Wolf}. Indeed a room temperature
magnetic semiconductor is likely essential for the development
of commercial spintronic devices, such as spin valves and transistors,
which exploit the fact that electrons have {\it spin} as well as charge. 
 
There has been similar recent activity in the field of ``multifunctional'',
or ``smart'' materials, which encompasses piezoelectrics, magnetostrictive
materials, shape memory alloys, and piezomagnetic materials, to name
a few. The progress has been driven in part by their application in 
micro-electromechanical and nano-electromechanical systems (MEMS and NEMS)
which are integrated micro-devices or systems combining electrical and 
mechanical components as sensors and actuators.
Piezoelectric materials are of particular interest, both because
of their fascinating fundamental physics and for their utility as
transducers between  electrical and mechanical energy. 
Applications are diverse and include medical ultrasound devices,
smart structures in automobiles, naval sonar and micromachines.

This paper describes a computational study of one possible avenue 
for {\it integration} of the fields of spintronics and smart materials:
the design of a piezoelectric semiconductor-based ferromagnet.
Many potential device applications can be envisaged for such a material
including  electric
field or stress-controlled ferromagnetic resonance devices, and variable
transducers with magnetically modulated piezoelectricity \cite{Wood_Austin}.
Also, the ability to couple with {\it either} the magnetic {\it or}
the electric polarization offers additional flexibility in the
design of conventional actuators, transducers and storage devices.
However the relationship between
ferromagnetism and piezoelectricity has not, to our knowledge, been
explored previously.

The material system that we focus on here is transition metal
doped ZnO, in which the transition metal is incorporated substitutionally
at the Zn site. This has been the subject of a number of
recent experimental studies focussed on the magnetic properties,
following a prediction using a simple mean field model that $p$-doped
samples should have a high ferromagnetic Curie temperature\cite{Dietl}.
While the most recent work on well-characterized (Zn,Co)O samples indicated
antiferromagnetic coupling between the Co ions\cite{Aditi} there have
also been a number of earlier reports of possible ferromagnetism. For
example Ueda \textit{et al.\/} \cite{Ueda} reported
that pulsed laser deposition (PLD) grown films of (Zn,Co)O with 
Co concentrations between 5 and 25 \% displayed
Curie temperatures between 280 K and 300 K, with a saturation
magnetization between 1.8 and 2.0 $\mu_{\rm B}$ per Co. They suggested that
in addition to possibile ``intrinsic'' ferromagnetism from
(Zn,Co)O, perhaps in the presence of hole carriers, 
the presence of cobalt oxide grains might also account for the observed
behavior.
Kim \textit{et al.\/} \cite{kim} also used PLD to grow (Zn,Co)O films and found
evidence for ferromagnetism when the films were grown under low O$_2$ partial
pressure. The origin of the ferromagnetism was not ``intrinsic'' however, but 
resulted from the formation of cobalt microclusters. Finally, Lim 
\textit{et al.\/} \cite{lim} reported ferromagnetism
in Zn$_{1-x}$Co$_x$O ($0.02 \le $x$ \le 0.40$) films grown by magnetron
co-sputtering on sapphire.  Other evidence for magnetic phenomena include 
observation of a large magnetic circular
dichroism signal at an energy corresponding to the ZnO band edge in
thin films of ZnO doped with Mn, Co, Ni and Cu \cite{Ando1,Ando2}.
This indicated significant influence of the transition metal on the ZnO states.
Also optical studies on laser ZnO:Co films (with some Al) on
sapphire confirmed that Co was divalent, high spin, and
substituting for Zn\cite{jin}. 

In addition to the proposed applications exploting the magnetic semiconducting
behavior, a ZnO-based
system has a number of other potential applications. First
it suggests the possibility of fabricating a {\it transparent} 
ferromagnet that will have great impact on industrial applications, for 
example in magneto-optical devices. Also, since the host ZnO 
material crystallizes in the wurtzite structure and is piezoelectric, 
we believe that transition metal doped ZnO is a promising system for 
designing the first ferromagnetic piezoelectrics. Piezoelectricity
in transition metal doped ZnO has not been explored to date.
                                                                          
In this paper we explore computationally the effects of doping a ZnO host 
with two transition metals (Mn and Co) over
a range of concentrations, and containing impurities that are
likely to occur in such systems. By calculating
the relative stability of the magnetic phases of the candidate materials,
we identify those with the strongest ferromagnetic tendencies,
and by inspecting the band structures we assess the most promising for
piezoelectricity. 
Previous authors have used density functional calculations to predict 
magnetic ground states in such systems. Sato and Katayama-Yoshida
\cite{Sato1} predicted that, within the Korringa-Kohn-Rostoker and local
spin density approximations, a ferromagnetic ground state would be favored
at 25\%-doping for most 3$d$ transition metals. For Mn-substitution 
they predicted that hole doping would induce ferromagnetism.
The same authors have recently extended their study to lower transition metal 
concentrations using the coherent potential approximation to model the 
dilute alloy, and reached the same conclusions \cite{Sato2}.
The main result of our work, however, is that ferromagnetism is in
general {\it not} strongly 
favored in ZnO doped with either Co or Mn, unless additional dopants
which provide $p$-type carriers are also present.

The remainder of this paper is organized as follows. In the next
Section we describe the technical details for the calculations 
performed in this work. In Section~\ref{FM_origin} we describe our
results for (Zn,Mn)O and (Zn,Co)O, as well as the intriguing
possibility of creating a ferromagnet using only vacancy doping.
Finally, in Section~\ref{Summary} we present our conclusions and
propose a practical scheme for creating a robust ZnO-based 
ferromagnet.

\section{Computational Details}
\label{details}

\subsection{SIESTA}

In order to simulate realistic dopant concentrations, we need to
perform calculations for supercells containing a large number
of atoms. Therefore we use a density functional theory (DFT) approach 
based on pseudopotentials with localized atomic orbital basis sets. 
This method, implemented
in the code SIESTA \cite{Siesta1,Siesta2,Siesta3},
combines accuracy with  small computational cost, particularly 
compared to other approaches such as plane wave pseudopotential
or all-electron methods. For a comprehensive description of the
SIESTA program, and its use in understanding the properties of
related systems, see in particular Refs.~\cite{Siesta2} and \cite{Stefano_2}.
Our pseudopotentials are designed using the standard Troullier-Martins 
formalism \cite{TM} with non-linear core corrections \cite{Lou82} 
and Kleinman-Bylander factorization \cite{KB1}. We use the Ceperley-Alder 
local spin density (LSDA) exchange-correlation functional \cite{CA}, and 
include scalar relativistic effects for Mn, Co and Zn. For Zn, we include 
the $(3d)^{10}$ electrons in the valence manifold, and construct the 
pseudopotential using a Zn$^{2+}$ reference configuration  with cut-off 
radii ($r_c$s) of 2.0, 2.1 and 1.9 a.u. for the $4s$, $4p$ and $3d$ 
orbitals respectively, and a partial core radius of 0.6 a.u. The 
eigenvalues and excitation energies of related configurations calculated 
from this pseudopotential agreed with the all-electron values to within 
4 mRy, and the bulk lattice constants, atomic positions and band structures 
(calculated using maximal basis sets) were in good agreement with all-electron 
values \cite{DalCorso}. We use the Mn pseudopotential of 
Refs.~\cite{Stefano_1} 
and \cite{Stefano_2}, which has been extensively tested in calculations for 
MnAs 
and (Ga,Mn)As. The reference configuration was 4$s^2$4$p^0$3$d^5$ with cut-off 
radii for the $s$, $p$ and $d$ components of  2.00, 2.20 and 1.90~au 
respectively. The Co pseudopotential, constructed from a
$4s^1 3d^8$ reference configuration, $r_c$s of 2.0 a.u. for all valence
orbitals and a partial core cut-off of 0.75 a.u., was previously
tested for Co metal and shown to give good agreement with all-electron
LSDA lattice constants\cite{Goto,Moroni}.  The oxygen pseudopotential 
was constructed from a $2s^2 2p^4$ 
reference configuration, $r_c$s of 1.15 a.u. for each angular momentum 
channel and no core corrections. Eigenvalues and excitation energies for 
related atomic states agreed within 1 mRy of the all-electron values. The
oxygen pseudopotential has also been tested extensively in calculations 
for bulk oxides\cite{Junquera_Ghosez}.

In localized orbital calculations, care must also be taken in
the optimization of the basis set.
The procedure to generate the numerical atomic orbital basis is
described in Ref.~\cite{Siesta4}. Several parameters determine the accuracy
of the basis, including the number of basis functions, the angular
momentum components included and the confinement radii. All these
have been optimized here to achieve the required accuracy.
The calculated structural and electronic properties of ZnO were largely 
unchanged on reducing from the maximal basis to two unpolarized basis 
functions ($\zeta$s) per orbital, therefore we
decided to use such a double-$\zeta$ basis for each Zn and O orbital.
Since we are primarily interested in the magnetic properties of diluted systems
which contain a small number of magnetic ions, we followed Ref.~\cite{Stefano_2}
in using a triple-$\zeta$ basis for the d orbitals on Co and Mn.
Note that we can afford to use triple-$\zeta$ for Co and Mn $d$ since there are
only a small number of these ions in the unit cell. In contrast the use of 
larger basis sets for Zn and O yields a more dramatic increase of the size 
of the computations. For the atomic orbital confinement radii 
we use the well-tested values from Ref. \cite{Stefano_2} of 6.0, 6.0 
and 5.0 a.u. for the Mn $s$, $p$ and $d$ orbitals. Use of the Mn values 
for Zn and Co, 
combined with 5.5 a.u. for O $s$ and $p$ orbitals, gave energy differences 
between ferro- and anti-ferromagnetic states that agreed well with those for 
larger confinement radii, while maintaining a reasonable computation time.

Other computational details include a 4 x 4 x 3 Monkhorst Pack grid for 
32 atom total energy calculations, with a 10 x 10 x 8 interpolation for 
density of states calculations, a real space mesh cut-off of 180 Ry and 
the neglect of non-overlap interactions.

\section{Origin and optimization of ferromagnetism in transition-metal doped 
ZnO.}
\label{FM_origin}

We begin by performing DFT calculations for bulk 
(Zn,Mn)O and (Zn,Co)O to search for the presence of ferromagnetism,
and to determine the nature of the interactions driving the magnetic
ordering. We choose Co as a prototypical dopant since the only 
current experimental report
of ferromagnetism in a ZnO-based system is in Co-doped ZnO films.
Mn is attractive for our theoretical studies, since
the half-filled $3d$ band of the Mn$^{2+}$ ion might lead to more
straightforward interactions in the (Zn,Mn)O band structure.

\subsection{(Zn,Mn)O}
\label{ZnMnO}

First we calculate the electronic properties of a 32 atom wurtzite structure
supercell of ZnO containing 1 Mn atom substituted at a Zn site. The supercell 
is formed by doubling the
conventional four atom wurtzite structure along each axis and the experimental
lattice constant is adopted, giving lattice vectors
\begin{center}
\begin{tabular}{r r r}
   0.500 & -0.866 &  0.0000 \\
   0.500 &  0.866 &  0.0000 \\
   0.000 &  0.000 &  1.6024  
\end{tabular}
\end{center}
and a lattice constant of 12.28 a.u.  Note that we do not perform a structural
optimization of the atomic positions, since the forces on the atoms in this ideal
structure are not large.
Since only 1 Mn atom is included in
each supercell the overall magnetic ordering is constrained to be ferromagnetic,
and the concentration of Mn ions is 6.25 \%. We obtain a calculated magnetic 
moment of 4.9 $\mu_B$ per unit cell, close to the value of 5.0 $\mu_B$
predicted for purely ionic Mn$^{2+}$ with five unpaired $3d$ electrons.

\begin{figure}[ht]
\centerline{\epsfig{figure=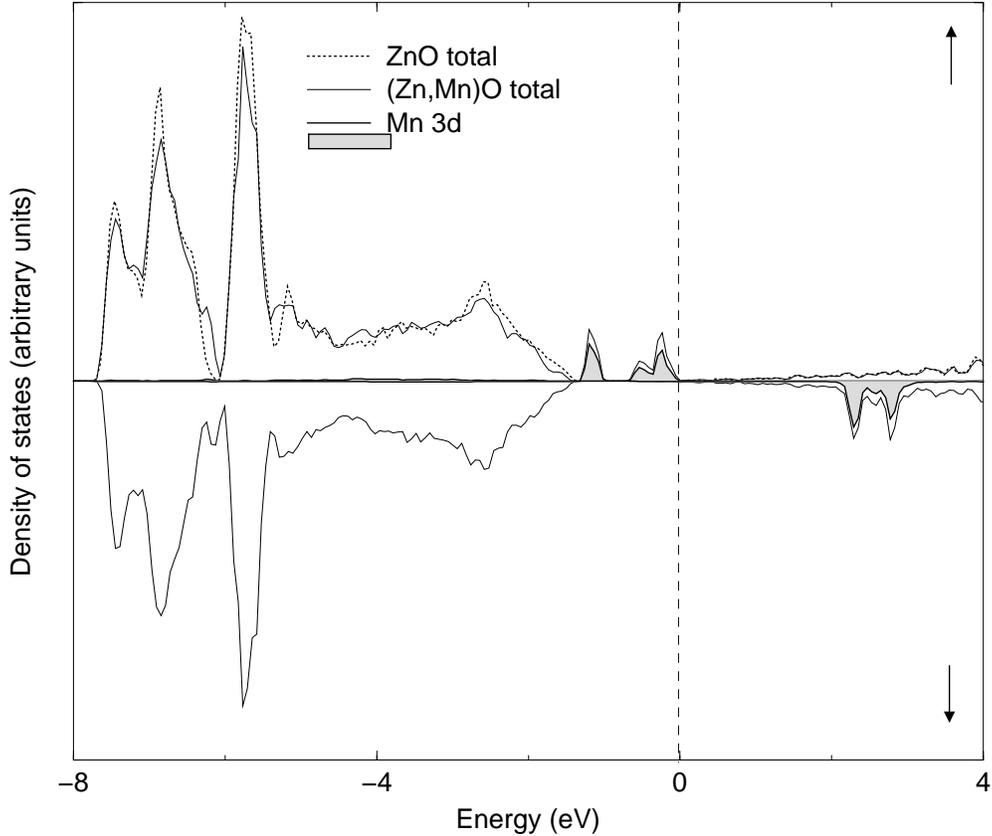,width=4.5in, angle=-90}}
\caption{Calculated density of states in ferromagnetic (Zn,Mn)O with a 
manganese dopant concentration of 6.25 \%. The solid line is the (Zn,Mn)O
total density of states and the gray shading indicates the contribution
from the Mn $3d$ orbitals. The positive $y$ axis shows the majority (up) spin
states and the negative $y$ axis the minority (down) spin states. The ZnO total
density of states is shown with a dashed line in the majority sector
for comparison.}
\label{ZnMnODOS}
\end{figure}

Our calculated total density of states for this system is shown as the 
thin solid line in Figure~\ref{ZnMnODOS}. The majority (up) spin 
states are plotted along the positive $y$ direction, 
and the minority (down) spin along negative $y$, with the pure ZnO density
of states shown as the dotted line in the $+y$ direction for comparison. 
The dashed line shows the energy of the highest occupied state, which we
have set to 0 eV. Note that the density of states of Mn-doped ZnO is largely
similar to that of undoped ZnO. 
The Zn $d$ states are concentrated within a largely unpolarized region
centered around -7 eV with a dispersion of about 2 eV, and the 
oxygen $p$ states are found in the region between -6 eV and -2 eV.
This is in good agreement with other LDA calculations\cite{DalCorso},
but with respect to experimental photoemission spectroscopy\cite{book}, 
the $3d$ bands 
calculated within LDA are too high in energy and overlap with the $sp^3$ 
valence band manifold. This causes an overestimation of the Zn $3d$ - O $2p$ 
hybridization \cite{Hill_Waghmare} which also shrinks the energy range of 
the $sp^3$ bands. The Mulliken population analyses indicate hybridization 
between the oxygen $2p$ electrons and the cations which is clearly
significant in spite of the overestimation. The Zn up and 
down spin occupations are each 5.45 per atom (they would be 5.0 for a purely 
ionic Zn$^{2+}$ ion) while those of O are 3.55 per atom (compared to the 
ionic value of 4.0), indicating electron transfer out of the O$^{2-}$ ion 
and onto the Zn ions. 

The Mn $d$ states (shaded gray, with a thick solid line) are largely
localized within the band gap region. The Mn $d$ manifold is split by 
an exchange splitting of around 4 eV into distinct groups of up-spin and 
down-spin states, each of which in turn is split by the smaller crystal 
field splitting (around 1 eV)
into the (lower energy) doublet  $e$ and higher energy triplet $t$ states. 
The up-spin Mn states form an impurity state just at the top of the valence 
band edge which is completely filled. The down-spin states are slightly 
broadened by overlap with the Zn $s$ states at the bottom of the conduction 
band. Both up- and down-spin Mn occupations are increased over their purely
ionic values (which would be 5.0 and 0.0 respectively) to 5.30 (due to 
occupation of the up-spin Mn $4s$ orbitals) and 0.75 (due to occupation 
of both the down spin $4s$ and $3d$.

Finally note that the LDA fundamental energy gap is underestimated by 
$\sim$ 40 \% compared with experiment.
This results in a non-zero density of states at the
Fermi energy, with the bottom of the Zn $4s$ conduction band overlapping 
slightly with the top of the up-spin Mn $3d$ states. Although this is
likely an artifact of the underestimation of the band gap by the LDA,
and in practice the system is likely to be insulating, it does preclude
the calculation of piezoelectric constants in this case.
However calculations for MnO in the wurtzite structure 
indicate a large piezoelectric response (around 30 \% larger than that
of ZnO) which augurs well for piezoelectricity in the mixed systems\cite{Priya}.

When two Mn atoms are included in a 32 atom unit cell (giving a 12.5 \%
doping concentration), a number of different magnetic and positional
arrangements are possible. Here we adopt two different pairs of Mn
positions; the `close' configuration, in which the Mn ions in the
same unit cell are separated by a single O ion, and the `separated'
configuration in which they are connected via -O-Zn-O- bonds. 
For both configurations we calculate the relative energies of the
ferromagnetic and antiferromagnetic spin alignments. Note that our
calculations show that the close configuration is more energetically favorable
than the separated one by around 10 meV, suggesting that the Mn ions
are likely to cluster together during growth, rather than distribute
themselves evenly throughout the lattice.

In all cases the energy differences between ferro- and antiferromagnetic
alignment are small (of the order of meV) with the antiferromagnetic state
being more favorable for the separated configuration and the ferromagnetic
being more favorable for the close configuration. Note that these energy
differences are so small that we believe that they are not significant within 
the uncertainties
of the DFT calculation. We therefore predict that, in the absence of
additional carriers or impurities, substitution of Zn by Mn ions in ZnO
will produce paramagnetic behavior down to low temperature. This is
in agreement with the earlier DFT calculations on this system\cite{Sato1,Sato2}.

Next we simulate $n$-type doping by removing an oxygen atom to form
an oxygen vacancy, as far as possible from the Mn ions. We find that
this stabilizes the antiferromagnetic state for both close and separated
configurations, but again by only one or two meV. In contrast,
the incorporation of a Zn vacancy,
representing $p$-type doping, stabilizes the ferromagnetic state of
both configurations by significant amounts - 
the separated configuration by around 10 meV, and the close configuration
by 60 meV - over the antiferromagnetic state. 
Notice that, in this case, there
are free carriers in the system, and so the
resulting metallicity will not allow piezoelectricity to occur.
In the next subsection we analyze in detail why the introduction of $p$-type
carriers stabilizes the ferromagnetic phase.

\subsection{(Zn,Co)O}

\begin{figure}[ht]
\centerline{\epsfig{figure=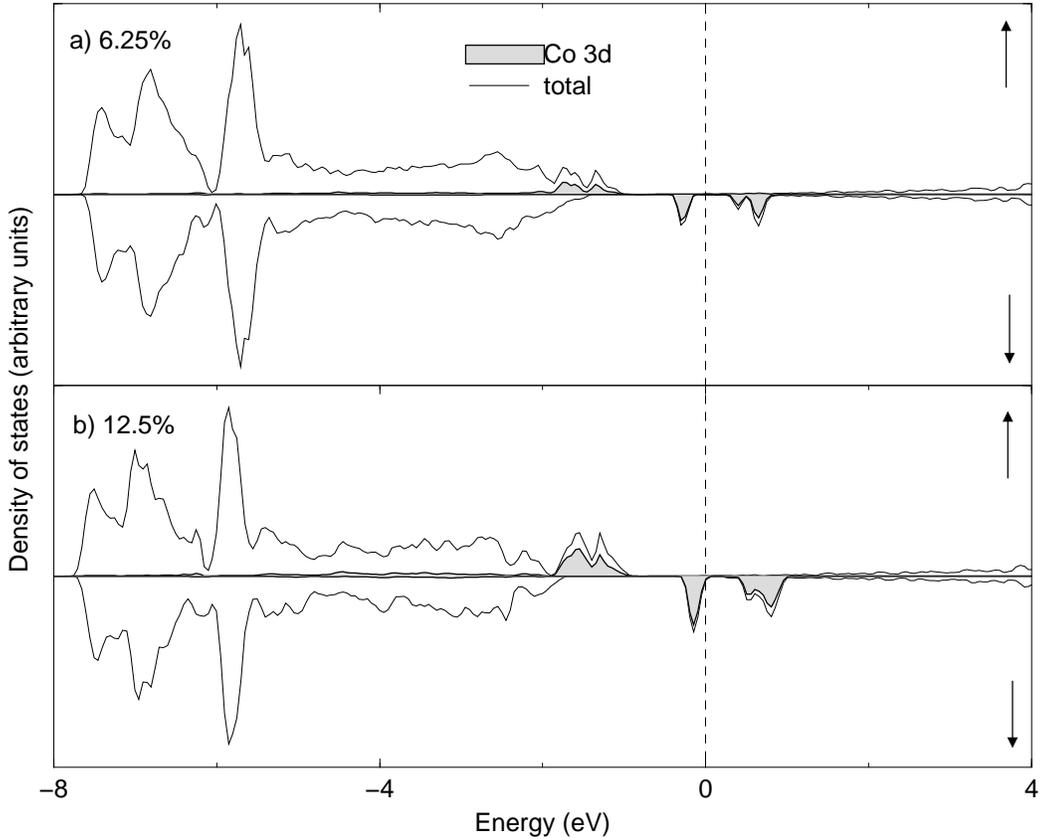,width=4.5in, angle=-90}}
\caption{Calculated densities of states in ferromagnetic (Zn,Co)O with 
cobalt dopant concentrations of (a) 6.25 \% and (b) 12.5 \%. The thin solid lines
show the (Zn,Co)O
total densities of states and the thick solid lines with the gray shading 
indicate the contribution from the Co $3d$ orbitals. Majority spin states
are plotted along the positive $y$ axis, and minority spin along  $-y$.}  
\label{ZnCoODOS}
\end{figure}

\begin{figure}[ht]
\centerline{\epsfig{figure=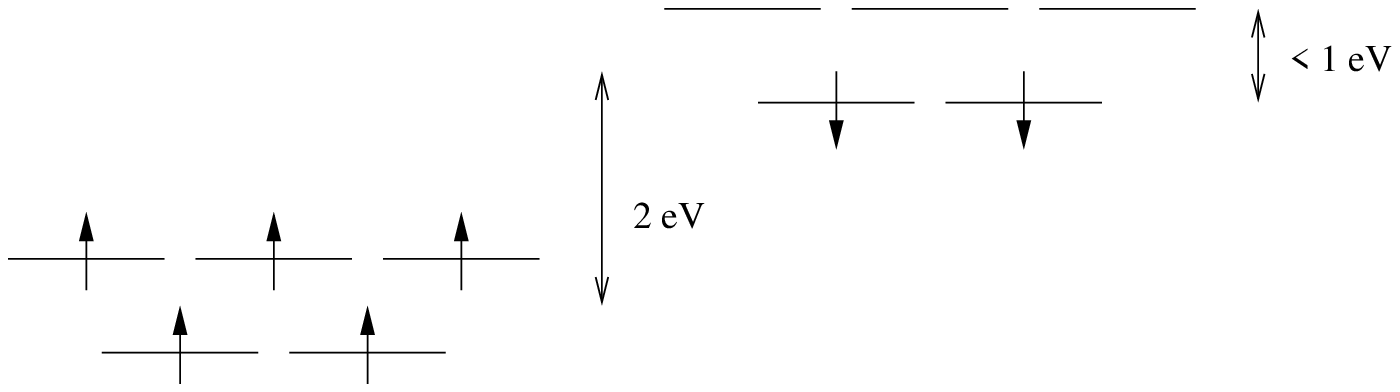,width=4.5in}}
\caption{Schematic energy diagram for a Co$^{2+}$ ion in a tetrahedral
oxide ligand field. The exchange splitting of around 2 eV is considerably
larger than the crystal field splitting (less than 1 eV).}
\label{cfvsex}
\end{figure}

Next we repeat the calculations described in Section~\ref{ZnMnO}, but
with Co instead of Mn as the magnetic dopant ion.
Again we begin with a 32 atom unit cell containing a single Co atom, and
consequently having ferromagnetic ordering. 
The density of states is shown in Figure~\ref{ZnCoODOS} (a). 
The majority spin Co $d$ states are lower in energy
than their Mn counterparts and so are more strongly hybridized 
with the O $2p$ states at the top of the valence band.
As in the case of Mn, the up-spin Co $d$-states are fully occupied.
However, in contrast to the formally $d^5$ Mn$^{2+}$, in the case of
formally $d^7$ Co$^{2+}$, the down-spin $d$ states are also
partially occupied, with the Fermi level lying in a gap between
the crystal field split $e$ and $t_2$ states.
This arrangement occurs because the exchange
splitting (between  up-spin ($\uparrow$)  and down-spin ($\downarrow$) states) is around
2 eV, while the crystal field splitting (between $e$ and $t_2$ states)
is much less than 1 eV. The $d^7$
configuration is therefore $e(\uparrow)^2, \ t_2(\uparrow)^3,
\ e(\downarrow)^2 \ t_2(\downarrow)^0$, as sketched in Figure~\ref{cfvsex}.
(Note that in the Mn-doped compound, the crystal field splitting
was even smaller (around 0.5 eV) and the exchange splitting larger (at
around 2.5 eV)).
The magnetic
moment per unit cell is calculated to be 3.10 $\mu_B$, close to the
value of 3.0 $\mu_B$ predicted for purely ionic Co$^{2+}$. 

When two Co atoms are included in a 32 atom unit cell (giving a 12.5\%
substituent concentration)
we find that, for both close and separated configurations, the energies of
the FM and AFM configurations are similar. This time, the FM is slightly
lower (the energy of the separated FM configuration is 4 meV lower
than that of the separated AFM separated configuration, and for the
close configuration the FM is only 1 meV lower than the AFM).
Again we believe that these small energy differences, which would anyway 
correspond
to Curie temperatures of the order of around 10 to 40 K, are not significant
within the uncertainties of the DFT calculation.
Our results are in sharp contrast to earlier calculations\cite{Sato1}
which predicted the ferromagnetic state to be 200 meV lower than the
antiferromagnetic in (Zn,Co)O, corresponding
to a ferromagnetic Curie temperature of around 2000 K.
 
The calculated density of states of the 12.5 \% doped ferromagnetic
system, with the Co atoms separated, is shown in Figure~\ref{ZnCoODOS} (b).
The structure is similar to that at 6.25 \% doping except for a broadening
of the Co bands as expected. The total magnetic moment per unit cell
is 6.1 $\mu_B$ for both close and separated configurations, again close to the
ideal ionic value of 3.0 $\mu_B$ per ion for high spin Co$^{2+}$.

Again
we have modeled hole doping in our system by removing a Zn atom from the 32
atom unit cell of 12.5 \% substituted (Zn,Co)O.
Such $p$-type doping strongly
stabilizes the FM state. In both the separated and close configurations 
the FM state is now 60 meV lower in energy than the AFM state.
The magnetic moment increases to 8 $\mu_B$ per unit cell (4 $\mu_B$ per Co), up from
the value of 6.1 $\mu_B$ for the un-doped case. This is consistent with
the removal of two minority spin electrons from each supercell. Note however
that Mulliken population analysis reveals that this additional magnetic 
moment in fact resides on the oxygen atoms rather than being localized on
the Co ions. The occupation of the Co $d$ states is 5 up-spin electrons
and 2.4 down-spin electrons in both ferro- and antiferromagnetic cases.
(Of course the AFM case has a complementary Co ion with 5 down-spin and
2.4 up-spin electrons.) It is also notable that the minority electrons
are distributed approximately evenly between all five $d$ orbitals.
This indicates that the interaction with the hole states causes 
strong ligand field effects which overcome to some extent the crystal
field splitting into occupied $e$ and empty $t_2$ minority states.
                                                
\begin{figure}[ht]
\centerline{\epsfig{figure=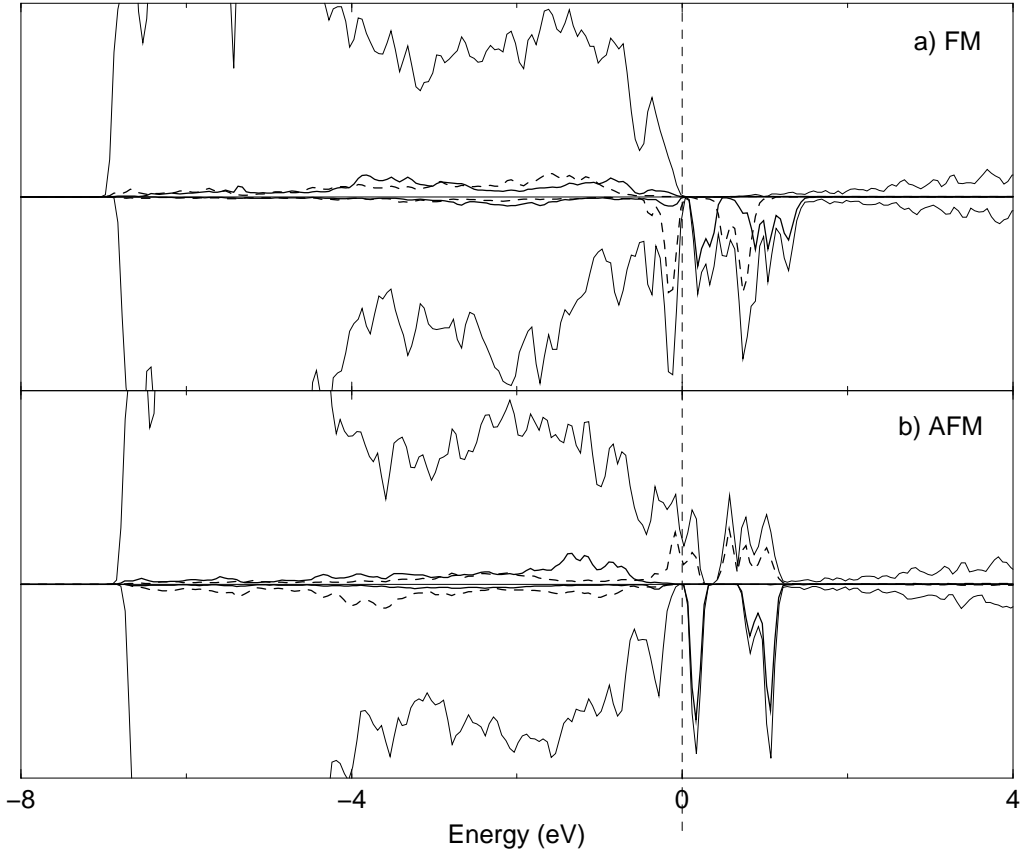,width=4.5in, angle=-90}}
\caption{Calculated density of states in a) ferromagnetic and b) 
antiferromagnetic (Zn,Co)O with a 
cobalt dopant concentration of 12.5 \% (Co ions in the separated 
configuration) and a single Zn vacancy.  The thin
solid line is the (Zn,Co)O total density of states and the thick black
solid and dashed lines indicate the contributions from the $3d$ orbitals
on each Co ion separately.}                     
\label{ZnCoO_FMZnf}
\end{figure}

The densities of states for both the ferro- and antiferromagnetic hole-doped
systems are shown in Figure~\ref{ZnCoO_FMZnf}. Note that the $y$-axis has been 
expanded compared with those in the earlier figures, so that the Co states occupy 
most of the range. The two Co ions
in each unit cell are plotted separately, one shown by a thick solid line and 
one by a thick dotted line. In both cases, as expected, the Fermi energy is 
shifted down
relative to the broad oxygen $2p$ band compared with the undoped cases.
(Since $E_F$ is set to 0 eV, this is manifested by a shift up in energy
of the bottom of the valence band). We see that the majority electrons
on each Co ion hybridize very strongly with the oxygen $p$ states, giving a
broad band with a width comparable to that of the O $p$ band. The increase
in hybridization compared with the undoped case is notable. Also
striking in comparison with the undoped case is the mixing of the
minority spin electrons with the oxygen band. Although the differences
between the FM and AFM electronic structures are indeed subtle, the
FM arrangement seems to allow for stronger energy-lowering hybridization
seen in the slightly broader bandwidth compared with the AFM case.
Also the FM arrangement allows a gap opening at the Fermi energy
which might contribute to its stabilization. Note however that the gap
is vanishingly small, and so any piezoelectric response would be
expected to be rather lossy.

In contrast, if the vacancy is created on the anion site (O atom replaced by
a $\Box$), the resulting $n$-doping stabilizes the AFM state. In the $n$-doped
separated configuration, the AFM state is now 4 meV lower in energy than the 
FM, and for the close configuration the AFM state is 1 meV lower than the FM.

\subsection{ZnO with vacancies}

Finally for this section, we investigate the intriguing possibility that
a ferromagnetic state could be obtained in a simple oxide solely from
interactions between vacancies\cite{Sawatzky}.  First we introduce a
single Zn vacancy into our 32 atom unit cell, and initialize our calculation
with the two oxygen atoms adjacent to the vacancy polarized either parallel 
or antiparallel to 
each other. The two starting conditions converge to different final
solutions with magnetizations of 1.8 $\mu_B$ and 0.4 $\mu_B$ per
supercell. A fully antiferromagnetic solution, with total magnetization of
0.0 $\mu_B$, is not obtained. The energy
of the more spin-polarized solution is lower than that of the less
polarized by 100 meV. Next we repeat the calculation with {\it two} Zn
vacancies. Again two spin-polarized solutions are obtained, one with a
magnetization of 2.9 $\mu_B$ per unit cell, and the other with 
0.6 $\mu_B$ per supercell, and again the more spin-polarized solution is
lower in energy by 100 meV per supercell. Although we cannot guarantee 
the absence of other (possibly AFM) minima, our results do suggest that
solutions with larger spin polarization have lower energy in such
vacancy-doped systems. Note however that the vacancy concentrations in
both cases studied here are much larger than realistic experimental values. 

\section{Summary and future work}
\label{Summary}

Our results suggest that, in contrast to earlier predictions, robust
ferromagnetism will only be obtained in transition metal doped ZnO if 
$p$-type carriers are also included. The method that we employed 
computationally, namely imposing a high concentration of Zn vacancies,
is clearly unfeasible experimentally. In fact, $p$-type ZnO is notoriously
difficult to realize, although there have been successes reported
using co-doping techniques \cite{Joseph}. In this final section we propose
that doping with Cu should be a feasible way of making ZnO $p$-type.
A relativistic Cu pseudopotential, generated from a $4s^1$ $3d^{10}$
ground state with $r_c$s of 2.1 a.u. for each orbital and partial 
non-linear core corrections was used. A double zeta plus polarization 
basis for $4s$ and $4p$ orbitals, and triple zeta plus polarization
for $3d$, with confinement radii of 7.0, 7.0 and 6.5 a.u., gave identical 
lattice constants,
bulk modulus and band structure for bulk Cu to those of plane wave and 
ultra-soft pseudopotential calculations. 

\begin{figure}[ht]
\centerline{\epsfig{figure=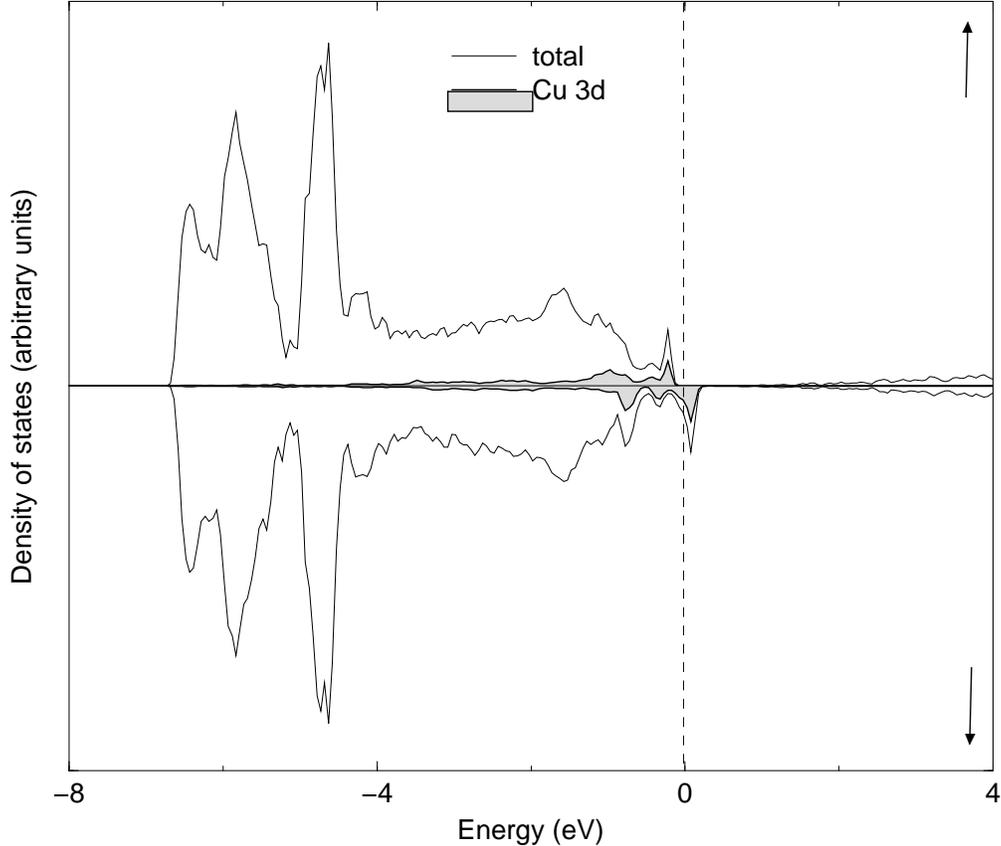,width=4.5in, angle=-90}}
\caption{Calculated density of states in (Zn,Cu)O with a 
copper dopant concentration of 6.25 \%.  The solid line is the (Zn,Cu)O
total density of states and the gray shading indicates the contribution
from the Cu $3d$ orbitals. Majority spin states are plotted along the
positive $y$ axis, and minority states along negative $y$. }     
\label{ZnCuODOS}
\end{figure}

We performed a total energy calculation for a 32 atom unit cell in which
one Zn atom was replaced by a Cu atom. Mulliken population analysis indicated
a total valence charge on the Cu atom of 10.3, compared to the ideal values
of 10 for Cu$^{+}$ or 9 for Cu$^{2+}$. This suggests that the Cu is 
adopting close to a +1 ionization state when it substitutes for a Zn$^{2+}$ ion,
indicative of $p$-type doping. Figure~\ref{ZnCuODOS} shows the 
total density of states for the system, with the Cu $3d$ 
partial density of states highlighted.  Note that the ``acceptor'' state 
above the Fermi energy is broader than
that seen in a protoptypical $p$-type semiconductor band structure because
the dopant concentration considered here is much higher.
It is clear that the up-spin
Cu states are completely filled, as expected, with the down-spin states
almost filled, giving a half-metallic band structure. As a result of the
half-metallicity, the magnetic moment corresponds to an integer number
of Bohr magnetons and is equal to 1 $\mu_B$. About half of the moment is
carried in the Cu $3d$ states (the Mulliken distribution of the Cu
charge between its atomic orbitals is 3$d_{\uparrow}^{5.0}$
3$d_{\downarrow}^{4.5}$ 4$s_{\uparrow}^{0.177}$ 4$s_{\downarrow}^{0.175}$), 
with the remainder delocalized across its
oxygen nearest neighbors. Note that these results were obtained within
the LSDA, and more detailed studies using beyond-LDA methods are ongoing.

Finally, we calculate the effect of simultaneously doping with both Co
and Cu, in the expectation that the $p$-type doping from the Cu will 
induce ferromagnetic interactions between the Co ions. We find this
indeed to be the case. When two separated Co ions are included in a 32 atom unit
cell with a single Cu ion (placed as far as possible from the Co ions) we obtain
a ferromagnetic ground state which is over 80 meV lower in energy than the
corresponding antiferromagnetic state. 
Note that there has been a recent experimental report of Cu-doping
inducing ferromagnetism with a Curie temperature of 550K in otherwise 
paramagnetic Fe-doped ZnO\cite{Han}.
The authors attribute the ferromagnetism to the presence of Cu$^{+}$ ions (observed
by x-ray absorption spectroscopy), although in this case 
Hall effect measurements indicate that the sample is strongly $n$-type.

In conclusion, our density functional theory calculations indicate that 
ferromagnetism can indeed be obtained by substituting Zn in ZnO with either
Co or Mn, but that it will only occur in the presence of simultaneous $p$-doping. 
We also propose a realistic scheme - the incorporation of Cu$^{+}$ ions on
the Zn sites - for achieving the $p$-doping. Although our results are 
encouraging for the production of ZnO-based ferromagnets, they do not
augur well for straightforward production of a ZnO-based magnetic
piezoelectric, since the additional carriers required to stabilize the
ferromagnetic state will be detrimental to the piezoelectricity.

\section{Acknowledgments}

NAS thanks David Clarke and Ram Seshadri for disucssions regarding suitable 
possible dopants for creating $p$-type ZnO.
This work was supported by the Department of Energy, grant number
DE-FG03-02ER45986, and made use of MRL central facilities, supported
by the the National Science Foundation under the
Award No. DMR00-80034. NAS thanks
the Earth Sciences Department at Cambridge University for their hospitality
during the sabbatical leave in which the calculations were performed.

\end{document}